\documentclass[aps,prd,twocolumns,eqsecnum,preprintnumbers,showpacs,amsmath,amssymb]
{revtex4}

\usepackage{graphicx}
\usepackage{dcolumn}
\usepackage{bm}

\begin{document}

\title{FAILURE OF THE LADDER APPROXIMATION TO QCD }

\author{V. Gogohia}
\email[]{gogohia@rmki.kfki.hu}

\affiliation{HAS, CRIP, RMKI, Depart. Theor. Phys., Budapest 114,
P.O.B. 49, H-1525, Hungary}

\date{\today}
\begin{abstract}
The proof of the failure of the ladder approximation to QCD is
given in manifestly gauge-invariant way. This proof is valid for
the full gluon propagator and for all types of quarks. The
summation of the ladder diagrams within the Schwinger-Dyson
integral equation for the quark-gluon vertex, on account of the
corresponding Slavnov-Taylor identity, provides an additional
constraint on the quark Schwinger-Dyson equation itself in the
ladder approximation. It requires that there is neither running
nor current quark masses in the ladder approximation. Thus, all
the results based on the nontrivial (analytical or numerical)
solutions to the quark Schwinger-Dyson equation in the ladder
approximation should be reconsidered, and its use in the whole
energy/momentum range should be abandoned.
\end{abstract}

\pacs{ 11.15.Tk, 12.38.Lg}

\keywords{}

\maketitle


\section{Introduction}

There are only three different types of interactions in Nature:
gravitational, electroweak and strong. However, all these
interactions are described by the gauge theories. They differ from
each other by the strength of the coupling constant, the gauge
group structure and the fundamental constituents involved. In the
two first theories the coupling constant is very weak and rather
weak, respectively. The gauge theory of strong interactions --
Quantum Chromodynamics (QCD) -- can be considered in both the
strong and the weak coupling regimes (asymptotic freedom)
\cite{1}. It is well known that the full dynamical information on
any gauge theory such as QCD and Quantum Electrodynamics (QED) is
contained in the corresponding quantum equations of motion, the
so-called Schwinger-Dyson (SD) equations for propagators (lower
Green's functions) and vertices (higher Green's functions)
\cite{1}. The Bethe-Salpeter (BS) type integral equations for
higher Green's functions and bound-state amplitudes \cite{2}
should also be included into this system. The Green's functions,
bound-state amplitudes and scattering kernels are the most
important objects in any gauge theory. This system should be
complemented by the corresponding Word-Takahashi (WT) identity in
QED and Slavnov-Taylor (ST) identities in QCD \cite{1}, which, in
general, relate lower and higher Green's functions to each other.
These identities are consequences of the exact gauge invariance
and therefore $"are \ exact \ constraints \ on \ any \ solution \
to \ QED/QCD"$ \cite{1}. To solve this system means to solve
QED/QCD itself and vice versa.

It is important to understand however, that the above mentioned
system is an infinite chain of strongly coupled highly nonlinear
integral equations, so there is no hope for an exact solution(s).
Let us also remind that this system is, in fact, an infinite chain
of relations between different propagators, vertices, kernels.
Thus, truncations/approximations are necessary for sure in order
to formulate a closed system of equations in this or that sector
of the theory. QED being a gauge theory with simple gauge group
$U(1)$, contains only two different sectors:
electron/positron-photon and BS in the sense that its system of
the dynamical equation of motion contains the electron/positron
and photon SD equations and the corresponding integral equation
for the electron/positron-photon vertex only. At the same time,
QCD being a gauge theory with rather complicated gauge group
$SU(3)$ color, contains many sectors of different nature, such as
quark, ghost, Yang-Mills (YM), Nambu-Goldstone (NG), BS, etc.
Making use of some truncation/approximation scheme in one sector,
it is necessary to be sure that nothing is going wrong in other
ones, since the SD system of equations may remain coupled even
after truncation/approximation made. As noted above, the main tool
to maintain the self-consistent treatment of different sectors in
QED/QCD is the use of the above-mentioned identities.

In theories with weak coupling constant the general method of
their investigation is the perturbation theory (PT). It allows one
to find solutions term by term in powers of the coupling constant.
In order to sum up the PT series just the above-mentioned system
of dynamical equations of motion is to be used. Within these
equations such kind of the summation is known as the ladder
approximation (LA) truncation scheme to QED and QCD (for its
general description see subsection below). Being thus a
generalization of PT, the main problem of the LA is, of course,
its self-consistency. There already exists one serious problem
with the LA to QCD, namely this truncation scheme is strongly
gauge dependent (and why this is not acceptable in QCD is
explained in Ref. \cite{3}). In the LA one should omit ghost and
non-Abelian corrections to the vertex and the corresponding
identity. They may appear in the full gluon propagator (see
below). By omitting ghosts, however, one cannot already cancel the
unphysical (longitudinal) degrees of freedom of the gauge bosons.
So, all results based on the LA will be, in principle, plagued by
the unphysical singularities (the unitarity of the $S$-matrix is
violated) in this case. This is the reason why the LA is usually
used in the Landau gauge only, in which the longitudinal component
of the full gluon propagator vanishes. On the other hand, this
means that any solution in the Landau gauge obtained within the LA
is a gauge artefact, and therefore cannot be used for the
calculation of any physical observable, which is, by definition,
manifestly gauge-invariant.

In Ref. \cite{4}, we have investigated the validity of the LA to
QCD, keeping the ghost degrees of freedom in the corresponding ST
identity for the quark-gluon vertex. It has been exactly proven
that even in this case the quark propagator is almost trivial one,
i.e., there is no "dressed" quark propagator in the LA to QCD.
This proof was given in manifestly gauge-invariant way and for any
types of quarks. We did not also use some specific solution for
the full gluon propagator, i.e., it remained completely arbitrary.
In this way, we have formulated an exact criterion to prove (in
QED) or disprove (in QCD) the LA in gauge theories. Nevertheless,
there are a lot of recent papers \cite{5,6,7} (and references
therein) in which the LA is being continuously used for the
calculation of physical observables in QCD.

The main purpose of this Letter is to explicitly show that the LA
to QCD is not only almost trivial one (which was proven in our
previous publication \cite{4}), but it does not take into account
the response of the true QCD vacuum at all (proven here), which is
only one modifies the quark self-energy (to make it "dressed").
The second point, which was also missed in the previous
publication \cite{4}, is to emphasize a possible connection
between the validity of the LA and the gauge group structure of
the corresponding theories (QED and QCD). Contrary to the previous
publication \cite{4}, the proof of the general failure of the LA
to QCD is given in such details which can be understood and
checked in all its steps by scientists who are not experts in this
field. Here we precisely describe how to formulate the
above-mentioned criterion (constraint) in the most transparent
way. We hope that all this will make it possible to prevent from
the future exploitation of the LA in QCD due to its internal
inconsistency proven in this Letter.

\subsection{General description of the LA}

Let us now describe what the LA is about in QED and QCD. In
general, it consists of approximating the full
quark/electron-gluon/photon vertices by their point-like ("bare")
counterparts in the corresponding kernels of the above-mentioned
SD integral equations for the quark/electron propagator and the
quark/electron-gluon/photon vertex (see Eq. (2.1) and Eq. (4.1)
below, respectively). Approximating in addition the full
gluon/photon propagator by its free expression in the BS-type
integral equation for the vertex, one gets the so-called quenched
LA (QLA) scheme, which became known as the "rainbow" approximation
for the quark/electron SD equation. However, very soon it has been
realized that QLA was too crude for QCD and should be improved in
order to incorporate the QCD renormalization group results for the
running coupling constant (asymptotic freedom) into the system of
the SD equations. The so-called improved LA (ILA) was proposed
\cite{8,9}. It makes it possible to take into account the
self-interacting gluon modes, including the 3-gluon coupling, at
least (non-Abelian character of QCD), at the level of the full
gluon propagator only. At the same time, the quark-gluon vertices
in the quark and in the kernel of the vertex SD equations remain
intact, i.e., they remain point-like ones (see below). Thus, the
exact definition of the LA excludes the non-Abelian corrections to
the quark-gluon vertices, while retaining them in the full gluon
propagator. The LA scheme with the full gluon propagator is known
as the generalized, or simply LA. For more detail description of
the LA in the quark and BS sectors see sections 2 and 3 below,
respectively.

\section{Quark SD equation}

In what follows we will investigate the validity of the LA by
comparing QED and QCD, and using precisely the corresponding
equations of motion. We begin with the SD equation for the quark
propagator. In the LA and in the momentum space it is

\begin{equation}
S^{-1}(p) = S^{-1}_0(p) + i \Sigma(p) = S^{-1}_0(p) - g^2_F \int
{d^4l \over {(2\pi)^4}} \gamma_\alpha S(l) \gamma_\beta D_{\alpha
\beta}(q),
\end{equation}
where $i\Sigma(p)$ is the quark self-energy and $q=p-l$. We assign
the factor $-ig \gamma$ to the point-like vertices with the
corresponding Dirac indices. Here $g^2_F = g^2 C_F$ and $C_F$ is
the eigenvalue of the quadratic Casimir operator in the
fundamental representation (for SU(N), in general, $C_F = (N^2 -
1)/2N = 4/3, \ N=3$ for QCD). The free quark propagator is

\begin{equation}
S^{-1}_0(p) = - i (\hat p - m_0)
\end{equation}
with $m_0$ being the current ("bare") mass of a single quark. Here
and everywhere below $D_{\alpha\beta}(q)$ is the full gluon
propagator in an arbitrary covariant gauge

\begin{equation}
D_{\alpha\beta}(q) = - i \left\{ \left[ g_{\alpha\beta} -
{{q_\alpha q_\beta}\over {q^2}} \right] d(q^2; \xi) + \xi
{{q_\alpha q_\beta}\over {q^2}} \right\}{1 \over q^2},
\end{equation}
and $\xi$ is the gauge fixing parameter (for example, $\xi=0$ --
Landau gauge). The free gluon propagator $D^0_{\alpha\beta}(q)$ is
given by Eq. (2.3) with $d(q^2; \xi) =1$, by definition. The
formal iteration solution in powers of the coupling constant
squared $g^2$ of the quark SD equation (2.1) with
$D_{\alpha\beta}(q)= D^0_{\alpha\beta}(q)$ is known as the
"rainbow" approximation, while with the full gluon propagator as
the generalized "rainbow" approximation. Just the latter one
includes all the non-Abelian gluon modes (the 3- and 4-gluon
couplings as well as the ghost-gluon vertex) at the level of the
full gluon propagator only (for the structure of the SD equation
for the full gluon propagator see Ref. \cite{1} and our paper
\cite{3}). As emphasized above, the non-Abelian corrections to the
quark-gluon vertices should be excluded, by definition, otherwise
the approximation becomes not "rainbow" at all. Let us note that
if in the full gluon propagator the special dependence on the
gluon momentum is chosen \cite{8,9}, then the corresponding
iteration solution is known as the improved "rainbow"
approximation, mentioned above.

Differentiating Eq. (2.1) with respect to $p_{\mu}$, one obtains

\begin{eqnarray}
\partial_{\mu} S^{-1}(p) =
- i \gamma_{\mu} + \partial_{\mu} i \Sigma(p)
 &=& -i \gamma_{\mu} - \partial_{\mu} g^2_F
\int {d^4l \over {(2\pi)^4}}
\gamma_\alpha  S(l) \gamma_\beta D_{\alpha \beta}(q) \nonumber\\
&=& -i \gamma_{\mu}  + g^2_F \int {d^4l \over {(2\pi)^4}}
\gamma_\alpha [\partial_{\mu} S(l)] \gamma_\beta D_{\alpha
\beta}(q),
\end{eqnarray}
up to an unimportant total derivative (as usual), which is assumed
to vanish at the ends of integration. This is the differential
form of the quark SD equation (2.1) relevant for further
discussion. Here and below $[\partial_{\mu} S(l)]$ denotes the
differentiation with respect to $l_{\mu}$. Let us remind that in
QED the electron SD equation (2.1) and hence its differential form
(2.4) is the same, apart from unimportant replacement $g^2_F
\rightarrow g^2$ for the case of QED.

Concluding, let us make a few remarks. In order not to complicate
notations, here and below we use the system of the SD equations
for the unrenormalized Green's functions. However, all the
integrals are divergent (this is the general feature of quantum
field theory, in particular QED and QCD). In order to assign a
mathematical meaning to them and formal (mainly algebraic)
operations with them, we will implicitly assume that they are
regularized at the upper limit, and the ultra-violet cut-off will
go to infinity at the final stage. Evidently, none of the exact
operations here and below will be affected by such assumed
regularization (for more remarks on the issue of renormalization
see below in section 7).

\section{ BS sector}

 It is well known that, in general, the SD equation for the
quark-gluon vertex contains four unknown scattering kernels
\cite{1}, from which only one is the BS scattering kernel, known
from QED. It is the sum of an infinite number of the corresponding
skeleton diagrams. The three others are rather complicated
objects, containing among them the ghost-quark scattering kernel.
The two others scattering kernels are to be combined with the 3-
and 4-gluon vertices. However, in order to proceed to the LA for
the quark-gluon vertex it is necessary to omit all the
above-mentioned three scattering kernels and retain only one,
namely the BS scattering kernel. That is why we call the
below-displayed integral equation as the BS type or simply BS
integral equation. In turn, only the first skeleton diagram should
be retained from this kernel. Moreover, the quark-gluon vertices
in this term should be replaced by their point-like counterparts.
Just in this way one gets the BS type integral equation in the LA
for the quark-gluon vertex $\Gamma_\mu^a(p,k)$, which analytically
can be written down as follows:

\begin{equation}
\Gamma_{\mu}^a (p,k) = -i \gamma_{\mu}T^a -  g^2 \int {d^4l \over
{(2\pi)^4}} \gamma_\alpha T^b S(l) \Gamma_{\mu}^a (l,k) S(l-k)
\gamma_\beta T^b D_{\alpha \beta}(q),
\end{equation}
where $k=p-p'$ is the momentum transfer and again $q=p-l$. In QED
the same integral equation takes place, apart from the $SU(3)$
color group generators, $T^a, \ T^b$. Just this group structure
makes the principal difference between QED and QCD (see below).
The coupling constant $g$ is already cancelled from both sides of
this equation. The formal iteration solution of this equation with
$D_{\alpha\beta}(q)= D^0_{\alpha\beta}(q)$ is known as the QLA,
mentioned above, while with the full gluon propagator as the
generalized LA. Let us emphasize once more that just the latter
one includes all the non-Abelian gluon modes (the 3- and 4-gluon
couplings as well as the ghost-gluon vertex) at the level of the
full gluon propagator only. As underlined above, the non-Abelian
corrections to the quark-gluon vertices should be excluded, by
definition, otherwise the approximation becomes not "ladder" at
all. If again in the full gluon propagator the special dependence
on the gluon momentum is chosen \cite{8,9}, then the corresponding
iteration solution is known as the ILA, also mentioned above.

For our purposes, it is sufficient to exactly decompose the vertex
and the quark propagator as follows:

\begin{eqnarray}
\Gamma_{\mu}^a (p,k) &=& \Gamma_{\mu}^a (p,k) + \Gamma_{\mu}^a
(p,0) - \Gamma_{\mu}^a (p,0) = \Gamma_{\mu}^a (p,0) + O_{\mu}^a
(p,k), \nonumber\\
 S(l-k) &=&  S(l-k)+ S(l)- S(l) = S(l) + O(l,k),
\end{eqnarray}
where, obviously, the terms $O_{\mu}^a (p,k)$ and $O(l,k)$ depend
on the momentum transfer $k$ linearly and higher, i.e., they are
terms, at least, of the order $k$. Here a few short remarks are in
order. Due to the correspondence between the point-like
quark-gluon vertex and the proper one, when all its momenta goes
to zero, the analyticity of the vertex at zero momentum transfer
in QCD is implicitly assumed. Non-analytical dependence (if it
makes sense at all) is completely different story, and is beyond
the scope of this paper. At the same time, how to remove
unphysical (kinematical) singularities from the vertex is
well-known procedure \cite{10}.

Substituting now these exact decompositions into the previous BS
equation, one obtains

\begin{equation}
\Gamma_{\mu}^a (p,0) + O_{\mu}^a(p,k) = -i \gamma_{\mu}T^a -  g^2
\int {d^4l \over {(2\pi)^4}} \gamma_\alpha T^b S(l)
[\Gamma_{\mu}^a (l,0) S(l) + L_{\mu}^a(l,k)] \gamma_\beta T^b
D_{\alpha \beta}(q),
\end{equation}
where

\begin{equation}
L_{\mu}^a (l,k) = \Gamma_{\mu}^a (l,0) O(l,k) + O_{\mu}^a(l,k)S(l)
+ O_{\mu}^a (l,k) O(l,k).
\end{equation}
The composition $L_{\mu}^a (l,k)$ depends on $k$ linearly, at
least. Equating now the terms at zero order in powers of the
momentum transfer $k$, the previous BS equation becomes equivalent
to the system of two equations, namely

\begin{equation}
\Gamma_{\mu}^a (p,0) = -i \gamma_{\mu}T^a -  g^2 \int {d^4l \over
{(2\pi)^4}} \gamma_\alpha T^b S(l) \Gamma_{\mu}^a (l,0) S(l)
\gamma_\beta T^b D_{\alpha \beta}(q),
\end{equation}
and

\begin{equation}
O_{\mu}^a(p,k) = -  g^2 \int {d^4l \over {(2\pi)^4}} \gamma_\alpha
T^b S(l) L_{\mu}^a(l,k) \gamma_\beta T^b D_{\alpha \beta}(q).
\end{equation}
The BS equation (3.5) will be the main subject of our
consideration. The BS equation (3.6) is, in fact, an infinite
chain of equations, obtained by equating the terms at every order
in powers of the momentum transfer $k$ (starting at the order
$k$). Each of these equations, however, should be solved on
account of the solution to the zero order equation (3.5), which
should be agreed with the solution of the WT identity (see below).

\section{ WT identity}

The ST identity in QCD contains the above-mentioned ghost-quark
scattering kernel \cite{1}. In order to get the SD equation in the
LA for the quark-gluon vertex it should be omitted (see discussion
above). So, omitting it here for the sake of self-consistency this
identity becomes the WT identity of QED and has the form

\begin{equation}
k_{\mu}\Gamma_{\mu}^a (p,p-k) = T^a S^{-1}(p) - S^{-1}(p-k)T^a.
\end{equation}
It provides an exact solution for the zero momentum transfer $k
=p-p'=0$ as ($\Gamma_{\mu}^a (p,p) \equiv \Gamma_{\mu}^a (p,0)$)

\begin{equation}
\Gamma_{\mu}^a (p,0) = T^a \partial_{\mu} S^{-1}(p).
\end{equation}
In what follows we will use the relation

\begin{equation}
\partial_{\mu} S^{-1}(p) = - S^{-1}(p)[\partial_{\mu} S(p)] S^{-1}(p),
\end{equation}
which stems from the obvious identity $S^{-1}(p)S(p)=
S(p)S^{-1}(p) =1$.

\section{Triviality of the quark propagator in the LA}

On account of the WT identity (4.2) and the relation (4.3), the BS
equation (3.5) becomes

\begin{equation}
\partial_{\mu} S^{-1}(p)T^a = -i \gamma_{\mu}T^a
+ g^2 T^b T^a T^b \int {d^4l \over {(2\pi)^4}} \gamma_\alpha
[\partial_{\mu} S(l)] \gamma_\beta D_{\alpha \beta}(q).
\end{equation}

Let us now use the commutation relation between color group
generators $[T^a, T^b] = if_{abc} T^c$, where $f_{abc}$ are the
antisymmetric $SU(3)$ structure constants with non-zero values,
given, for example in Ref. \cite{11}. After doing some group
algebra, one obtains

\begin{equation}
T^b T^a T^b = [C_F -{1 \over 2}C_A] T^a,
\end{equation}
where $C_F$ is the above mentioned eigenvalue of the quadratic
Casimir operator in the fundamental representation while $C_A$ is
the same but in the adjoint representation (for $SU(N)$, in
general, $C_A = N =3$ for QCD). So, from this relation and because
of the notation $g^2 C_F = g_F^2$, one arrives at

\begin{equation}
\partial_{\mu} S^{-1}(p) = -i \gamma_{\mu}
+ [g^2_F - {1 \over 2} g^2 C_A]  \int {d^4l \over {(2\pi)^4}}
\gamma_\alpha [\partial_{\mu} S(l)] \gamma_\beta D_{\alpha
\beta}(q),
\end{equation}
where we have already cancelled the color group generator $T^a$
from both sides of this equation (In QED the factor $[g^2_F - (1
/2) g^2 C_A]$ is to be simply replaced by $g^2$). Comparing Eq.
(5.3) with that of Eq. (2.4) (the second line), one immediately
concludes that

\begin{equation}
 - {1 \over 2} g^2 C_A  \int {d^4l \over {(2\pi)^4}}
\gamma_\alpha [\partial_{\mu} S(l)] \gamma_\beta D_{\alpha
\beta}(q) =0.
\end{equation}
However, comparing the first and second lines in Eq. (2.4), and
taking into account the last relation, one obtains that it can be
reduced to

\begin{equation}
\partial_{\mu} i \Sigma(p) = 0.
\end{equation}
 This constraint has only a trivial solution, namely

\begin{equation}
\Sigma(p) = m_c,
\end{equation}
where $m_c$ is the constant of the dimensions of mass (constant of
integration). From the quark SD equation (2.1), taking into
account Eq. (2.2), and on account of the trivial solution to the
quark self-energy (5.6) obtained above, it follows that

\begin{equation}
S(p) = { i  \over \hat p  - (m_0 + m_c) }.
\end{equation}
 Thus, the solution of the constraint equation (5.6) in the LA to
covariant gauge QCD requires the quark propagator to be a free
one, apart from the redefinition of the quark mass. In other
words, there should be neither analytical nor numerical solution
to the quark SD equation in the LA. So, there is $no \
"running/dressed"$ quark mass in the LA to QCD. However, the real
situation is even more catastrophic.

\section{Internal inconsistency of the LA to QCD}

 Substituting the obtained
solution for the quark propagator (5.7) back to the quark SD
equation (2.1), and taking the trace from its both sides and
omitting some rather simple algebra, one obtains

\begin{equation}
m_c= (m_0 + m_c)f(p^2)
\end{equation}
with

\begin{equation}
f(p^2) = i g^2_F \int {d^4l \over {(2\pi)^4}} {[ 3 d(q^2; \xi) +
\xi ] \over (l^2 - (m_0 + m_c)^2) q^2},
\end{equation}
where let us remind $q=p-l$ and the dependence on $\tilde{m}=m_0 +
m_c$ and $\xi$ is not shown, for simplicity. Since we are seeking
the solution of the relation (6.1) explicitly, the regularization
of the divergent integral (6.2) with the help of the assumed
ultra-violet cut-off is not appropriate. Obviously, the solution
in this case may depend on it which is not acceptable. In this
case, the regularization with the help of the corresponding
subtraction is much more appropriate. So, let us define as usual
$f_R(p^2) = f(p^2) - f(0)$, and the relation (6.1) should be then
replaced as

\begin{equation}
m_c= (m_0 + m_c)f_R(p^2).
\end{equation}
It does not depend on the ultra-violet cut-off and the finite
function enters it. Thus, its only solution is

\begin{equation}
m_c= m_0=0.
\end{equation}
 From Eq. (5.7) then it follows that the quark
propagator in the LA to QCD finally becomes

\begin{equation}
S(p) =  i / \hat p.
\end{equation}
This solution is not acceptable, since even the current masses of
light quarks are not, in general, zero. So, the LA to QCD cannot
be used, it is simply wrong. Let us underline that we do not
specify the full gluon form factor $d(q^2; \xi)$ and hence the
full gluon propagator itself.

Due to the expression (5.7),  the vertex at zero momentum transfer
is always trivial one anyway, i.e., it is

\begin{equation}
\Gamma_{\mu}(p,0) = -i \gamma_{\mu},
\end{equation}
while the BS integral equation (3.6) for the non-zero momentum
transfer is not trivial.

\section{Conclusions and discussion}

1. The proof of the general failure of the LA to QCD has been
given in manifestly gauge invariant way (without specifying the
gauge fixing parameter in the full gluon propagator), and for any
gluon propagator (the full, QLA, ILA or something else, it does
not matter), and for all types of quarks (light or heavy).

2. The failure of the generalized LA itself, stems from the color
charge interactions in QCD (nontrivial gauge group structure).

3. In theories without colors, for example in QED (simplistic
gauge group structure), there is no constraint. Formally, in this
case we can put $C_A=N = 0$ in Eq. (5.4), while omitting the
dependence on $N$ in the quark SD equation, making it thus the
electron SD equation. So, the constraint equation (5.4) will
identically vanish. In its turn this means that the electron SD
equation obtained from the BS equation, on account of the WT
identity as described in this Letter, completely coincides with
the electron SD equation itself (more precisely with its
differential form). This justifies the use of the LA in QED.

4. In QED an expansion in powers of the external momentum makes
not much sense, since there are no stable bound states (the
positronium is unstable).

5. In QCD the same expansion is relevant because of the existence
of the NG sector. Many important physical quantities such as
scattering length, pion charge radius, etc. are defined as
coefficients of the pion form factor expansion in powers of the
external momentum (the chiral perturbation theory \cite{12} and
its counterpart at the fundamental quark level \cite{13}). In QCD
the zero external momentum transfer has a physical meaning, as
relating directly to the various physical observables.

6. The self-consistency of any other truncation schemes, for
example such as planar, $1/N_c$ limit, etc. ought to be
investigated in the same or other way. As emphasized in our paper,
this is important in theories with complicated gauge group such as
QCD.

We have investigated the self-consistency of the LA to QCD and
QED. The wide-spread opinion that different sectors in QCD in this
approximation can be decoupled from each other is not justified.
They remain connected even in the LA (nontrivial gauge group
structure). This clearly shows the general failure of the LA to
QCD. All the nontrivial (analytical or numerical) results,
obtained by the solution of the quark SD equation only in the LA,
should be abandoned. The use of the LA to QCD in the whole
energy/momentum range is forbidden, while to use it in the high
energy/momentum region only (i.e., as a part of some another
approximation) is possible. Within the LA the non-Abelian degrees
of freedom (for example, the 3- and 4-gluon couplings) can be only
taken into account in the full gluon propagator, and such kind of
corrections to the quark-gluon vertices are forbidden, by
definition. So, there is no room for improvement in order to
formulate more sensible LA, i.e., it is too rigid truncation
scheme. In other words, it is either LA or it is completely
different approximation if one includes some other corrections to
the vertices, which investigation is completely beyond the scope
of this Letter, indeed.

Apparently, the smallness of the coupling constant in the gauge
theories is not sufficient to use PT (and the LA as its
generalization). This could be only first necessary condition. The
second sufficient condition is the structure of the corresponding
gauge group. It should be simple enough to valid the use of PT.
Thus, the use of PT in the theory of gravitation becomes doubtful,
since its gauge group is not so simple. One can change the regime
of the consideration in QCD (weak or strong couplings), but
impossible to change its gauge group structure. So, a possible
correspondence between gauge group structure and the complexity of
the corresponding vacuum structure should be also mentioned.
Simplistic gauge group structure corresponds to the almost simple
vacuum structure like in QED. Complicated gauge group structure
corresponds to the highly nontrivial structure of the true QCD
vacuum. In this connection let us remind that just its response,
which is only one which modifies the quark self-energy. In the LA
to QCD it cannot modify it, that is why it is simply wrong. Thus,
due to its internal inconsistency, we propose to completely
abandon it from the use in QCD.

Renormalization, of course, cannot change our main result, namely
the proof of the internal inconsistency of the LA. That is why we
started, for simplicity, from the unrenormalized Green's
functions. The issue of renormalization has been discussed in all
details (i.e., explicitly and carefully) in our previous
publication \cite{4}. It causes only technical complications,
which make no any sense to repeat here again. Evidently, the
internal inconsistency of the LA does not depend on the issue of
multiplicative renormalizability of QCD.

Ghost degrees of freedom cannot change this result either. In
principle, one may keep them in the quark-gluon ST identity,
making it "exact" in the LA to the above-mentioned ghost-quark
scattering kernel, while necessarily omitting them in the vertex
SD equation. This technical complication has been also
investigated in our previous publication \cite{4}. Since we do not
specify the full gluon propagator, the quark propagator in the LA
will remain an almost trivial one in any noncovariant gauge as
well.

\end{document}